\documentstyle[12pt,epsf]{ioplppt}
\def\Tc{T_{\rm C}}
\def\uu{^{\mbox{ }}}
\def\s{\sigma}
\def\bs{\mbox{\boldmath$\s$}}
\def\u{\uparrow}
\def\a{\alpha}

\def\dd{\downarrow}

\def\en{\epsilon}
\def\las{\langle}
\def\ras{\rangle}
\def\la{\left\las}
\def\lla{\la\la}
\def\llas{\las\las}
\def\ra{\right\ras}
\def\rra{\ra\ra}
\def\rras{\ras\ras}
\def\k{{\bf{k}}}

\def\S{{\bf{S}}}

\def\gmub{g\mu_{\rm B}}
\def\gmubB{\gmub B}
\def\bW{\bar W}
\def\bmu{\bar\mu}
\def\kB{k_{\rm B}}

\begin{document}\jl{3}

\title{Magnetic susceptibility of the double exchange model}
\author{ACM Green and DM Edwards}
\address{Department of Mathematics, Imperial College, London SW7 2BZ, UK}

\begin{abstract}
Previously a many-body coherent potential approximation (CPA) was
used to study the double exchange (DE) model with quantum local spins $S$,
both for $S=1/2$ and for general $S$ in the paramagnetic state.
This approximation, exact in the atomic limit, was considered to be a
many-body extension of Kubo's one-electron dynamical CPA for the DE model.
We now extend our CPA treatment to the case of general $S$ and spin
polarization. We show that Kubo's one-electron CPA is always recovered in
the empty-band limit and that our CPA is equivalent to dynamical mean
field theory in the classical spin limit. We then solve our CPA equations
self-consistently to obtain the static
magnetic susceptibility $\chi$ in the strong-coupling limit. As in the case
of the CPA for the Hubbard model we find unphysical behaviour in $\chi$ at
half-filling and no magnetic transition for any finite $S$. We identify the
reason for this failure of our approximation and propose a modification
which gives the correct Curie-law behaviour of $\chi$ at half-filling and
a transition to ferromagnetism for all $S$.
\end{abstract}


\section{Introduction}\label{sintro}

Recently there has been much interest in the perovskite manganite compounds
T$_{1-x}$D$_{x}$MnO$_3$ where T and D are trivalent and divalent cations
respectively. These exhibit a rich variety of phases including charge,
orbital, ferromagnetic and antiferromagnetic ordering
\cite{rRamirez,rCoey}. Of particular interest is La$_{1-x}$Ca$_x$MnO$_3$
with $x\sim 0.3$; in this compound ferromagnetic-paramagnetic and
metal-insulator transitions occur together, and for temperatures near the
critical temperature an applied magnetic field causes a very large
reduction in electrical resistance: this is the phenomenon known as
colossal magnetoresistance (CMR).

The physically relevant electrons in the manganites are those from the Mn
3d levels, which are split by the approximately cubic crystal field into
triply degenerate $t_{2g}$ levels and higher energy doubly degenerate
$e_g$ levels. Occupied $e_g$ levels are further split into two
nondegenerate levels by the Jahn-Teller effect. Electrons from the $e_g$
levels are able to hop between Mn sites via the O atoms, forming a narrow
conduction band, but those from the $t_{2g}$ levels are localised. There is
a strong Hund's rule coupling on the Mn sites, so the
$t_{2g}$ electrons are usually modelled as local $S=3/2$ spins
ferromagnetically coupled to the itinerant $e_g$ electrons. At $x=0$ there
is one $e_g$ electron per site so the system is a Mott insulator. Doping by
$D$ atoms produces holes in the $e_g$ band which enables conduction to
occur.

The simplest model for the CMR compounds, which neglects the $e_g$
degeneracy and any coupling to phonon modes, is Zener's \cite{rZener}
double exchange (DE) model
\begin{equation}
H=\sum_{ij\s}t_{ij}c^{\dag}_{i\s}c_{j\s}\uu-
J\sum_i\S_i\cdot\bs_i-h\sum_i L_i^z, \label{eHDE}
\end{equation}
where $i$ and $j$ are Mn sites, $c_{j\s}\uu$ ($c^{\dag}_{i\s}$) is a
$\s$-spin conduction electron annihilation (creation) operator,
${\S}_i$ is a local spin operator, $\bs_i$ is a conduction electron spin
operator, $L_i^z=S_i^z+\s^z_i$ is the $z$-component of the total angular
momentum on a site, $t_{ij}$ is the hopping integral with discrete Fourier 
transform $t_{\k}$, $J>0$ is the Hund's rule coupling constant, and
$h=\gmubB$ is the Zeeman coupling strength, $B$ being the applied magnetic
field.
The number of conduction electrons per atom $n$ is assumed to be given by
$n=1-x$. The idea of the DE model is that hopping of $e_g$ electrons
between neighbouring sites is easier if the local spins on the sites are
parallel, so an effective ferromagnetic coupling between the local spins is
induced by the conduction electrons lowering their kinetic energy. Double
exchange coupling differs from conventional Heisenberg coupling by being
(for classical local spins) of the form $\cos(\theta_{ij}/2)$ rather than
$\cos(\theta_{ij})$, where $\theta_{ij}$ is the angle between the
$i$- and $j$-site local spins.

According to Millis \etal \cite{rMillis1, rMillis2} the CMR effect arises
from a
competition between double exchange coupling, which produces a tendency
towards the conducting ferromagnetic state, and strong coupling to
phonon modes, which tends to localise the electrons via self-trapping.
In a previous paper \cite{rEdwardsGreenKubo} we confirmed that the DE
model above cannot account for the very high resistivity of the
paramagnetic state. Moreover
experiments show that coupling to the crystal
lattice is important \cite{rZhao, rBillinge, rTeresa}. 
In this paper however we will complete our study of the simple DE model,
aiming to understand the purely electronic properties of CMR systems
modelled by \eref{eHDE} before tackling more realistic and
complicated models. We concentrate particularly on the magnetic properties.

In \cite{rEdwardsGreenKubo} we derived an approximation
for the one-electron Green function which was based on Hubbard's scattering
correction approximation for the Hubbard model \cite{rHubbard}. In the
Hubbard model this
approximation is derived by decoupling the Green function equations of
motion according to an alloy analogy in which electrons of one spin are
frozen whilst the Green function for those of the opposite spin is
calculated.
For the finite $S$ DE model this approach is complicated by the possibility
of dynamic spin scattering--- conduction electrons exchanging angular
momentum with the local electrons. We obtained an approximation which,
like Hubbard's, was exact in the atomic limit for all band filling.
Since Hubbard's approximation is equivalent to the coherent potential
approximation (CPA) in the alloy analogy for the Hubbard model, and our
approximation
reduces to a one-electron dynamical CPA due to Kubo \cite{rKubo} in the
empty-band limit of the DE model we regard our approximation as a many-body
extension of the CPA.

In \cite{rEdwardsGreenKubo} we concentrated on simple cases in which the
CPA system of equations of motion closed easily, and calculated the
electronic structure and resistivity of the paramagnetic state. In
\sref{sCPAsoln} we formulate and solve the general CPA equations.
In \sref{sCPAsus} we then calculate the static magnetic susceptibility
self-consistently within the CPA. In 
\sref{sDMFT} we compare our CPA in the classical spin limit with dynamical
mean field theory, and in \sref{sDMFTTc} the CPA is modified so as to
improve the behaviour of the susceptibility.
A summary and outlook are given in \sref{ssummary}.

\section{Solution of the CPA equations}\label{sCPAsoln}

In this section we will use equation of motion (EOM) decoupling 
approximations to derive an expression for the one-electron Green
function $G^{ij}_{\s}=\llas c_{i\s}\uu\,;c^{\dag}_{j\s}\rras$ of the
DE model. In a previous paper \cite{rEdwardsGreenKubo} we obtained
$G^{ij}_{\s}$ in the special cases of zero-field paramagnetism and
saturated ferromagnetism for all values of $S$, but only considered the
case of arbitrary magnetisation for $S=1/2$. Here we extend this previous
treatment to the case of general $S$ and spin polarization. The
decoupling approximations used are direct extensions of those
used in \cite{rEdwardsGreenKubo}, which were in turn generalisations
of Hubbard's scattering correction approximation \cite{rHubbard} for
the Hubbard model.

We split the Green functions into components $G^{ij}_{\s}=
G^{ij+}_{\s}+G^{ij-}_{\s}$ where $G^{ij\alpha}_{\s}$ describes
propagation via singly ($\alpha=-$) and doubly ($\alpha=+$) occupied
sites: $G^{ij\alpha}_{\s}=
\llas n^{\alpha}_i c_{i\s}\uu\,;c^{\dag}_{j\s}\rras$ where
$n^+_i=n_i\uu$, $n^-_i=1-n_i\uu$, $n^+_{i\s}=n_{i\s}\uu$, and
$n^-_{i\s}=1-n_{i\s}\uu$. Here $n_{i\s}\uu=c^{\dag}_{i\s}c_{i\s}\uu$ and
$n_i=n_{i\u}+n_{i\dd}$.
Now the cases considered in
\cite{rEdwardsGreenKubo} were chosen so that the system of EOM could
be closed using only the Green functions $G^{ij\alpha}_{\s}$,
$\llas S^z_i n^{\alpha}_i c_{i\s}\uu\,;c^{\dag}_{j\s}\rras$, and
$\llas S^{-\s}_i n^{\alpha}_i c_{i-\s}\uu\,;c^{\dag}_{j\s}\rras$ where
$S^{-\s}_i=S^-_i$ and $S^+_i$ for $\s=\u,\dd$ respectively;
here we will also need the Green functions 
$\llas (S^z_i)^m n^{\alpha}_i c_{i\s}\uu\,;c^{\dag}_{j\s}\rras$ and
$\llas (S^z_i)^{m-1}S^{-\s}_i n^{\alpha}_i c_{i-\s}\uu\,;
c^{\dag}_{j\s}\rras$ for $m>1$ in order to close the system. It is
simplest to work with the Green functions
\numparts
\begin{eqnarray} \label{eSdef}
S^{ij\alpha}_{\s}(\lambda)=\lla \e^{\lambda S^z_i} n^{\alpha}_i
c_{i\s}\uu\,;c^{\dag}_{j\s}\rra\\
T^{ij\alpha}_{\s}(\lambda)=
\lla \e^{\lambda S^z_i}S^{-\s}_i n^{\alpha}_i c_{i-\s}\uu\,;
c^{\dag}_{j\s}\rra,
\end{eqnarray}
\endnumparts
which are in generating function form so that
$\partial^m S^{ij\alpha}_{\s}/\partial\lambda^m\vert_{\lambda=0}=
\llas (S^z_i)^m n^{\alpha}_i c_{i\s}\uu\,;c^{\dag}_{j\s}\rras$ and
$\partial^{m-1} T^{ij\alpha}_{\s}/\partial\lambda^{m-1}\vert_{\lambda=0}=
\llas (S^z_i)^{m-1} S^-_i n^{\alpha}_i c_{i\dd}\uu\,;c^{\dag}_{j\s}\rras$.

For notational simplicity we will work for $\s=\u$; the $\s=\dd$ EOM
can be obtained by making the replacements $c_{i\s}\mapsto c_{i-\s}$,
$c^{\dag}_{i\s}\mapsto c^{\dag}_{i-\s}$,
$S^z_i\mapsto -S^z_i$, $S^{\pm}_i\mapsto S^{\mp}_i$,
$h\mapsto -h$ and $\lambda\mapsto -\lambda$.
Recall that with the fermionic definition of Green
functions,
$\llas A\,;C\rras_{\en}=-i\int_0^{\infty}\d t\,\exp(\i\en t)\las\{A(t),C\}
\ras$, the EOM is
\begin{equation}
\en\,\lla A\,;C\rra_{\en}=\la\{A,C\}\ra+\lla[A,H]\,;C\rra_{\en}.
\end{equation}
We use the fact that
$S^{ij\alpha}_{\s}(\lambda)$ and $T^{ij\alpha}_{\s}(\lambda)$ are in
generating function form to write their exact EOM in the form
\begin{eqnarray}
\fl\nonumber
\left(\en+\frac{h}{2}+\frac{J}{2}\frac{\partial}{\partial\lambda}
\right)S^{ij\alpha}_{\u}(\lambda,\en)+
\frac{J}{2}\e^{\lambda\delta_{\alpha+}}T^{ij\alpha}_{\u}(\lambda,\en)=
\delta_{ij}\la \e^{\lambda S^z}n^{\alpha}_{\dd}\ra\\
+\sum_k t_{ik} \lla \e^{\lambda S^z_i}n^{\alpha}_{i\dd}
c_{k\u}\uu\,;c^{\dag}_{j\u}\rra_{\en}+\lla \e^{\lambda S^z_i}
\left[n^{\alpha}_{i\dd},H_0\right]c_{i\u}\uu\,;c^{\dag}_{j\u}
\rra_{\en}\label{ecpa1}
\end{eqnarray}
for $S^{ij\alpha}_{\u}(\lambda,\en)$ and
\begin{eqnarray}
\fl\nonumber
\left(\en+\frac{h}{2}-\frac{J}{2}\left(\delta_{\alpha-}+
\frac{\partial}{\partial\lambda}\right)\right)
T^{ij\alpha}_{\u}(\lambda,\en)+\frac{J}{2}
\e^{-\lambda\delta_{\alpha+}}\left(S(S+1)+
\alpha\frac{\partial}{\partial\lambda}-
\frac{\partial^2}{\partial\lambda^2}\right)
S^{ij\alpha}_{\u}(\lambda,\en)=\\
\nonumber
-\alpha\delta_{ij}\la\e^{\lambda S^z}S^-\s^+\ra+\sum_k t_{ik}
\lla \e^{\lambda S^z_i} S^-_i n^{\alpha}_{i\u} c_{k\dd}\uu\,;
c^{\dag}_{j\u}\rra_{\en}\\
+\lla \e^{\lambda S^z_i} S^-_i\left[n^{\alpha}_{i\u},
H_0\right]c_{i\dd}\uu\,; c^{\dag}_{j\u}\rra_{\en}\label{ecpa2}
\end{eqnarray}
for $T^{ij\alpha}_{\u}(\lambda,\en)$. Here $H_0=\sum_{ij\s}t_{ij}
c^{\dag}_{i\s}c_{j\s}\uu$ is the kinetic part of the Hamiltonian
and we have dropped the site indices in the expectations,
assuming the system to be in a homogeneous phase (i.e.\ we will not
consider antiferromagnetism). If one works directly with EOM for
the Green functions 
$\llas (S^z_i)^m n^{\alpha}_i c_{i\s}\uu\,;c^{\dag}_{j\s}\rras$ and
$\llas (S^z_i)^{m-1} S^-_i n^{\alpha}_i c_{i\dd}\uu\,;c^{\dag}_{j\s}\rras$,
as in \cite{rEdwardsGreenKubo}, one is faced with $4S+1$ algebraic
equations for general $S$. Here, using the generating functions, these
algebraic equations are reduced to just two differential equations where
differentiation with respect to $\lambda$ corresponds to the
coupling between the different algebraic equations.

These EOM form a closed system
apart from the undetermined Green functions on the right-hand sides
which correspond to the effects of hopping. The decoupling procedure
entails making approximations for these kinetic terms which close the 
system of equations; since these terms are proportional to $t$ this
procedure is exact in the atomic limit $t_{ij}\rightarrow 0$.
As mentioned in the introduction the idea of the alloy analogy
is to neglect the effects of the kinetic part $H_0$
of the Hamiltonian on electrons of one spin whilst considering the
propagation of an electron of the opposite spin. Accordingly we
neglect the final terms of \eref{ecpa1} and \eref{ecpa2} since the
occupation
number operators in these Green
functions are considered to be frozen in time. It remains to find
closed approximations for $\sum_k t_{ik} \llas
\exp(\lambda S^z_i)n^{\alpha}_{i\dd}c_{k\u}\uu\,;c^{\dag}_{j\u}
\rras_{\en}$ and $\sum_k t_{ik}
\llas \exp(\lambda S^z_i) S^-_i n^{\alpha}_{i\u} c_{k\dd}\uu\,;
c^{\dag}_{j\u}\rras_{\en}$. The derivation of scattering correction
approximations for these terms
is identical to that of \cite{rEdwardsGreenKubo} apart from the
occurrence of the factor $\exp(\lambda S^z_i)$ and will not be
repeated here. In fact we use
\numparts
\begin{eqnarray}
\fl\nonumber
\sum_k t_{ik} \lla \e^{\lambda S^z_i} n^{\alpha}_{i\dd}
c_{k\u}\uu\,;c^{\dag}_{j\u}\rra_{\en}\approx
\la \e^{\lambda S^z}n^{\alpha}_{\dd}\ra \sum_k t_{ik}
\lla c_{k\u}\uu\,;c^{\dag}_{j\u}\rra_{\en}\\ \label{eapprox1}
+J_{\u}(\en)\lla\left(\e^{\lambda S^z_i}
n^{\alpha}_{i\dd}-\la\e^{\lambda S^z}n^{\alpha}_{\dd}\ra\right)
c_{i\u}\uu\,;c^{\dag}_{j\u}\rra_{\en}\\ \label{eapprox2}
=\la \e^{\lambda S^z}n^{\alpha}_{\dd}\ra\left(
\sum_k t_{ik}G^{kj}_{\u}(\en)-J_{\u}(\en)G^{ij}_{\u}(\en)\right)
+J_{\u}(\en)S^{ij\a}_{\u}(\lambda,\en)\\
\fl\nonumber
\sum_k t_{ik}\lla \e^{\lambda S^z_i} S^-_i n^{\alpha}_{i\u}
c_{k\dd}\uu\,;c^{\dag}_{j\u}\rra_{\en}\approx
-\alpha\la\e^{\lambda S^z}S^-\s^+\ra\sum_k t_{ik}
G^{kj}_{\u}(\en)\\ \label{eapprox3}
+J_{\dd}(\en+h)
\lla\e^{\lambda S^z_i} S^-_i n^{\alpha}_{i\u}c_{i\dd}\uu\,;
c^{\dag}_{j\u}\rra_{\en}
+J_{\u}(\en)\lla\alpha\la\e^{\lambda S^z}S^-\s^+\ra
c_{i\u}\uu\,;c^{\dag}_{j\u}\rra_{\en}\\ \label{eapprox4}
=-\alpha\la\e^{\lambda S^z}S^-\s^+\ra\left(
\sum_k t_{ik}G^{kj}_{\u}(\en)-J_{\u}(\en)G^{ij}_{\u}(\en)\right)
+J_{\dd}(\en+h)T^{ij\a}_{\u}(\lambda,\en)
\end{eqnarray}
\endnumparts
where $J_{\s}(\en)=\en-\Sigma_{\s}(\en)-G_{\s}(\en)^{-1}$.
Here $\Sigma_{\s}(\en)$ is the self-energy, local within this
approximation, and $G_{\s}(\en)=G^{ii}_{\s}(\en)$ is the local
component of the Green function.
$J_{\s}(\en)$ contains the effects of coherent propagation of the
electron as a $\s$-spin from site $i$ back to site $i$ via paths
avoiding the site at intermediate stages \cite{rShiba}. Note that if an
$\u$-spin electron of energy $\en$ becomes a $\dd$-spin by
exchanging angular momentum with a local
spin it must then propagate at energy $\en+h$,
hence the occurrence of $J_{\dd}(\en+h)$ in \eref{eapprox3}
and \eref{eapprox4} above.
It may be seen that approximations \eref{eapprox1} and \eref{eapprox3}
close the system of equations \eref{ecpa1} and \eref{ecpa2}.
No further approximations are made.

For convenience we now define $E_{\s}(\en)=\en-J_{\s}(\en)$,
which will later be related to the Weiss function of
dynamical mean field theory \cite{rDMFT},
$E^h_{\s}(\en)=E_{\s}(\en+h\delta_{\s\dd})+\s h/2$, which puts the energy
shift effects of the magnetic field into $E_{\s}$, and
\begin{equation}
\lambda^{ij}_{\s}(\en)=\delta_{ij}+\sum_k t_{ik} G^{kj}_{\s}(\en)
-J_{\s}(\en)G^{ij}_{\s}(\en).\label{elambda}
\end{equation}
We make the above approximations so that \eref{ecpa1} and \eref{ecpa2}
become
\begin{eqnarray}\fl\nonumber
\left[\begin{array}{cc}
E_{\u}^h(\en)+J/2\,\frac{\partial}{\partial\lambda} &
J/2\,\e^{\lambda\delta_{\alpha +}}\\
J/2\,\e^{-\lambda\delta_{\alpha +}}[S(S+1)+\alpha
\frac{\partial}{\partial\lambda}-
\frac{\partial^2}{\partial\lambda^2}] &
E_{\dd}^h(\en)-J/2\,(\delta_{\alpha -}+
\frac{\partial}{\partial\lambda})
\end{array}\right]
\left(\begin{array}{c}
S^{ij\alpha}_{\u}(\lambda,\en)\\
T^{ij\alpha}_{\u}(\lambda,\en)
\end{array}\right)\approx\\
\lambda^{ij}_{\u}(\en)
\left(\begin{array}{c}
\las \e^{\lambda S^z}n^{\alpha}_{\dd}\ras \\
-\alpha\las\e^{\lambda S^z}S^-\s^+\ras
\end{array}\right).
\end{eqnarray}
This is a coupled pair of linear
differential equations (with respect to $\lambda$) of first and
second order respectively. The first equation is used to eliminate
$T^{ij\alpha}_{\u}(\lambda,\en)$ in terms of
$S^{ij\alpha}_{\u}(\lambda,\en)$,
\begin{equation}
\fl T^{ij\alpha}_{\u}(\lambda,\en)=\frac{2}{J}
\e^{-\lambda\delta_{\alpha+}}\left(\lambda^{ij}_{\u}(\en)
\la \e^{\lambda S^z}n^{\alpha}_{\dd}\ra-E^h_{\u}(\en)
S^{ij\alpha}_{\u}(\lambda,\en)-
\frac{J}{2}\frac{\partial}{\partial\lambda}
S^{ij\alpha}_{\u}(\lambda,\en)\right).\label{ecpa3}
\end{equation}
Substituting into the second equation we obtain a first order equation for 
$S^{ij\alpha}_{\u}(\lambda,\en)$,
\begin{eqnarray}
\fl\nonumber
\frac{\partial}{\partial\lambda}S^{ij\alpha}_{\u}(\lambda,\en)+
\left(
\frac{J/2\,S(S+1)-E_{\u}^h(\en)(2/J\,E^h_{\dd}(\en)+\alpha)}
{E^h_{\u}(\en)-E^h_{\dd}(\en)}\right)
S^{ij\alpha}_{\u}(\lambda,\en)=\\
-\la\frac{(2/J\,E_{\dd}^h(\en)+
\alpha-S^z)\e^{\lambda S^z}n^{\alpha}_{\dd}
+\alpha\,\e^{\lambda(\delta_{\alpha+}+S^z)}
S^-\s^+}{E^h_{\u}(\en)-E^h_{\dd}(\en)}\ra
\lambda^{ij}_{\u}(\en).\label{ecpa4}
\end{eqnarray}

This equation is of the form
\begin{equation}
\frac{\partial S(\lambda)}{\partial\lambda}+P\,S(\lambda)=R(\lambda),
\end{equation}
where $P$ is independent of $\lambda$, which has the solution
\begin{equation}
S(\lambda)=C\,\e^{-P\lambda}+\e^{-P\lambda}\int^{\lambda}\d\lambda\,
\e^{P\lambda}R(\lambda)
\end{equation}
where $C$ is a constant of integration. Furthermore by 
inserting local spin projection operators into definition \eref{eSdef}
of $S^{ij\alpha}_{\u}(\lambda,\en)$ it may be seen that 
$S^{ij\alpha}_{\u}(\lambda,\en)=\sum_{m=-S}^S a^{ij\alpha}_{m\u}(\en)
\exp(m\lambda)$ where $a^{ij\alpha}_{m\u}(\en)$ is independent of
$\lambda$.
Since in general $P$ is not an integer we must have $C=0$. Hence we find
\begin{eqnarray}
\fl\nonumber
S^{ij\alpha}_{\u}(\lambda,\en)=
\la\frac{\Bigl(E^h_{\dd}(\en)+J/2(\alpha-S^z)\Bigr)
\e^{\lambda S^z}n^{\alpha}_{\dd}-(J/2\,\delta_{\alpha-})
\e^{\lambda S^z}S^-\s^+}
{\Bigl(E^h_{\u}(\en)+JS^z/2\Bigr)\Bigl(E^h_{\dd}(\en)+J/2(\alpha-S^z)
\Bigr)-J^2/4(S+\alpha S^z)(S+1-\alpha S^z)}
\right.\\
\left.\hskip -0.75in
+\frac{(J/2\,\delta_{\alpha+})\e^{\lambda(S^z+1)}S^-\s^+}
{\Bigl(E_{\u}^h(\en)+J/2(S^z+1)\Bigr)\Bigl(E^h_{\dd}(\en)-JS^z/2\Bigr)
-J^2/4(S-S^z)(S+1+S^z)}
\ra\lambda^{ij}_{\u}(\en)\label{eS}
\end{eqnarray}
where we adopt the convention that a quotient of operators,
$B$ divided by $A$, means $A^{-1}B$.
Substituting into \eref{ecpa3} we also obtain
\begin{eqnarray}
\fl\nonumber
T^{ij\alpha}_{\u}(\lambda,\en)=
\la
\frac{\e^{\lambda(S^z-\delta_{\alpha+})}\Bigl(
J/2(S^z+\alpha S)(S^z-\alpha(S+1))
n^{\alpha}_{\dd}+\delta_{\alpha-}(E^h_{\u}(\en)+JS^z/2)S^-\s^+\Bigr)}
{\Bigl(E^h_{\u}(\en)+JS^z/2\Bigr)\Bigl(E_{\dd}^h(\en)+J/2(
\alpha-S^z)\Bigr)-J^2/4(S+\alpha S^z)(S+1-\alpha S^z)}\right.\\
\hskip -0.75in
\left.-\frac{\delta_{\alpha+}\Bigl(E^h_{\u}(\en)+J/2(1+S^z)\Bigr)
\e^{\lambda S^z}S^-\s^+}{\Bigl(E^h_{\u}(\en)+J/2(1+S^z)\Bigr)
\Bigl(E_{\dd}^h(\en)-JS^z/2\Bigr)-J^2/4(S-S^z)
(S+1+S^z)}\ra\!\lambda^{ij}_{\u}(\en).\label{eT}
\end{eqnarray}
It is easy to check that these expressions reduce to those obtained in
\cite{rEdwardsGreenKubo} in the appropriate cases.

We are mostly interested in the local components of the Green functions,
$S^{\a}_{\u}=S^{ii\a}_{\u}$ and $T^{\a}_{\u}=T^{ii\a}_{\u}$, and in
the following we drop site indices. Note that
$\lambda^{ii}_{\s}(\en)=1$; this follows from the relation
$\lambda^{ij}_{\s}(\en)=G^{ij}_{\s}(\en)/G_{\s}(\en)$ which is easy to
obtain from the definition \eref{elambda} of $\lambda^{ij}_{\s}(\en)$
using Fourier transforms and the locality of $\Sigma_{\s}(\en)$, as in
section $3.1$ of \cite{rEdwardsGreenKubo}.

Since $E^h_{\s}$ is a functional of $G_{\s}$, \eref{eS} with $i=j$
determines $G_{\s}$, in principle, in terms of the expectations 
$\las\exp(\lambda S^z)\ras$, $\las\exp(\lambda S^z)n_{\s}\ras$, and
$\las\exp(\lambda S^z)S^{\mp}\s^{\pm}\ras$ as functions of $\lambda$.
These expectations must be evaluated self-consistently in terms of
$S^{\a}_{\s}(\lambda)$ and $T^{\a}_{\s}(\lambda)$. The equation for
$G_{\s}$
simplifies if we approximate the physical (simple cubic tight-binding) bare
density of states (DOS) by an elliptic DOS,
$D(\en)=2\sqrt{W^2-\en^2}/(\pi W)$ where $W$ is the half-bandwidth; for
this
DOS it may be shown that $E_{\s}(\en)=\en-W^2/4\,G_{\s}(\en)$ so that
$E_{\s}$ is an explicit function of $G_{\s}$. 
We now define the functional
\begin{equation}
{\rm I}[g]=\oint_{\gamma}\frac{\d\en}{2\pi i}f(\en-\mu)\,g(\en)
\end{equation}
where $f$ is the Fermi function, $\mu$ is the chemical potential, and
$\gamma$ is the anticlockwise contour lying just below and just above the
real axis. This functional is useful owing to the usual sum rule
\begin{equation}
{\rm I}\left[\,\lla A\,;C\rra\,\right]=\la CA\ra,\label{esumrule}
\end{equation}
from which it follows (exactly) that
\numparts
\begin{eqnarray}
I\left[S^{\a}_{\s}(\lambda)\right]=\la\e^{\lambda S^z}n^{\a}_{-\s}
n_{\s}\uu\ra\label{esumruleS}\\
I\left[T^{\a}_{\s}(\lambda)\right]=\la\e^{\lambda S^z}S^{-\s}
\s^{\s}\ra\delta_{\a-}.\label{esumruleT}
\end{eqnarray}
\endnumparts
The expectations $\las\exp(\lambda S^z)n_{\s}\ras$ and
$\las\exp(\lambda S^z)S^{\mp}\s^{\pm}\ras$
can therefore be obtained directly
from the sum rule as $\las\exp(\lambda S^z)n_{\s}\ras=I[S_{\s}(\lambda)]$,
where $S_{\s}(\lambda)=\sum_{\alpha}S^{\alpha}_{\s}(\lambda)$,
and $\las\exp(\lambda S^z)S^{-}\s^{+}\ras=I[T_{\u}(\lambda)]$. However 
$\las\exp(\lambda S^z)\ras$ must be obtained indirectly, in principle,
by solving the
system of equations obtained by applying the I-functional to \eref{eS}
and \eref{eT}. Knowledge of $\las\exp(\lambda S^z)\ras$ is equivalent to
knowledge of $P(S^z)$, the probability distribution function for local
spins, and it is not clear that this quantity will be accurately obtained
from the (approximate) single-electron Green functions that we have
considered here. In fact, as will be seen later, the self-consistent
determination of this quantity causes problems with our CPA.

We now specialise to three important cases of particular interest: $n=0$,
$J=\infty$, and $S=\infty$, in which \eref{eS} and \eref{eT} simplify
considerably. The results below generalise previous work
\cite{rEdwardsGreenKubo} which was restricted, for general magnetisation,
to the case $S=1/2$.

\subsection{The empty band limit}

In \cite{rKubo} Kubo used a one-electron dynamical CPA to derive an
expression for $G_{\u}$ valid in the low-density limit $n\rightarrow 0$.
From \eref{eS} with $\lambda=0$ we calculate $G_{\u}\uu$ in this limit as
\begin{equation}
\fl
G_{\u}=\la
\frac{E^h_{\dd}-J/2\left(1+S^z\right)}
{\left(E^h_{\u}+J/2\,S^z\right)\left(E^h_{\dd}-J/2\left(1+S^z\right)\right)
-J^2/4\left(S+1+S^z\right)\left(S-S^z\right)}\ra.
\end{equation}
This is equivalent to Kubo's equation for $G_{\u}$ so our
decoupling approximation is indeed a many-body extension of the CPA.

\subsection{The strong coupling limit}

In the physical systems for which the double exchange model was introduced
$J\gg t_{ij}$ and $0<n<1$. In this situation the chemical potential lies
in the lowest band near $-JS/2$,
so we shift the energy origin, $E^h_{\s}\mapsto E^h_{\s}-JS/2$, and let
$J\rightarrow\infty$. Equations \eref{eS} and \eref{eT} then become
\numparts
\begin{eqnarray}
S^{\a}_{\u}(\lambda)=
\la\e^{\lambda S^z}\frac{\left(S+1+S^z\right)n^-_{\dd}
+S^-\s^+}{\left(S+1+S^z\right)E^h_{\u}+\left(S-S^z\right)E^h_{\dd}}\ra
\delta_{\a-}\label{eSsc}\\
T^{\a}_{\u}(\lambda)=
\la\e^{\lambda S^z}(S-S^z)\frac{\left(S+1+S^z\right)n^-_{\dd}
+S^-\s^+}{\left(S+1+S^z\right)E^h_{\u}+\left(S-S^z\right)E^h_{\dd}}\ra
\delta_{\a-}.\label{eTsc}
\end{eqnarray}
\endnumparts

\subsection{The classical spin limit}

In dynamical mean field theory (DMFT), a local approximation exact in the
infinite dimensional limit \cite{rDMFT}, the double exchange model can be
solved exactly in the classical spin limit $S\rightarrow\infty$ in which
the local spins can be treated as static \cite{rFurukawa}.
Since our approximation is also local
it is interesting to compare our results with DMFT in this limit. We let
$S\rightarrow\infty$ in \eref{eS} and \eref{eT}, 
scaling $J$, $\lambda$, $h$ and $T_{\s}^{\alpha}(\lambda)$
as $1/S$, and obtain
\numparts
\begin{eqnarray}
S^{\a}_{\u}(\lambda)=
\la\e^{\lambda S^z}\frac{(E_{\dd}-J/2\,S^z)n^{\a}_{\dd}
+\a J/2\, S^-\s^+}{(E_{\u}+J/2\,S^z)(E_{\dd}-J/2\,S^z)
-J^2/4(1-(S^z)^2)}\ra \label{eSSinf}\\
T^{\a}_{\u}(\lambda)= \la\e^{\lambda S^z}
\frac{J/2((S^z)^2-1)n^{\a}_{\dd}-
\a(E_{\u}+J/2\, S^z)S^-\s^+}
{(E_{\u}+J/2\,S^z)(E_{\dd}-J/2\,S^z)
-J^2/4(1-(S^z)^2)}\ra \label{eTSinf}
\end{eqnarray}
\endnumparts
where by $S^z$ and $S^-$ we mean $S^z/S$ and $S^-/S$ respectively.
Note that in this limit $E^h_{\s}=E_{\s}$.
In \sref{sDMFT} below we will derive these Green functions within DMFT for
comparison, and it will be found that our CPA agrees with DMFT for
$S^{\a}_{\s}$ and $T_{\s}\uu$, but not for $T^{\pm}_{\s}$.

\section{Self-consistent CPA susceptibility}
\label{sCPAsus}

In this section we calculate the static magnetic susceptibility $\chi$ of
the zero-field paramagnetic state. For simplicity we specialise to the
strong coupling limit $J=\infty$ most favourable to ferromagnetism and
use the elliptic bare DOS mentioned in the previous section. We drop the
spin suffices on zero-field paramagnetic state quantities and define
$\delta A$ to be the first order deviation of any quantity $A$ from
its value in this state. Thus $\delta A$ is proportional to the applied
magnetic field $B$ or equivalently to $\delta\las L^z\ras=
\delta\las S^z+\s^z\ras$. We proceed by calculating
$\delta\las L^z\ras$ in terms of the Zeeman energy $h=\gmubB$
and using $\chi=(\gmub)^2\lim_{h\rightarrow 0}(\delta\las L^z\ras/h)$.

We first derive a couple of useful identities. Equations
\eref{eSsc} and \eref{eTsc} imply that $T_{\s}(\lambda)=
(S-\partial/\partial\lambda)S_{\s}(\lambda)$. We apply the sum rule 
\eref{esumrule}--\eref{esumruleT} to this relation and, using the fact that
all the self-consistently determined expectations are real, obtain
the rather obvious results
\numparts
\begin{eqnarray}
\la \e^{\lambda S^z}\S\cdot\bs\ra=S/2\, \la \e^{\lambda S^z}n\ra
\label{eid1}\\
\la S^z n\ra = 2S\la\s^z\ra.\label{eid2}
\end{eqnarray}
\endnumparts
These serve as a check on our approximation and will later be used to
manipulate expectations.

The Green functions are especially simple in the ($J=\infty$)
zero-field paramagnetic state: from \eref{eSsc} with $\lambda=0$ and
$\alpha=-$ we have
\begin{equation}
G(\en)=\frac{(S+1-n/2)/(2S+1)}{E(\en)},
\end{equation}
which corresponds to a band of weight $(S+1-n/2)/(2S+1)$ per spin. 
In the elliptic DOS case, for which $E(\en)=\en-W^2/4\,G(\en)$,
we can easily solve this equation to obtain
\begin{equation}
G(\en)=\frac{2}{W^2}\left(\en-\sqrt{\en^2-\bW^2}\,\right)\label{eGpara}
\end{equation}
where $\bW=W\sqrt{(S+1-n/2)/(2S+1)}$ is the half-width of the renormalised
band, which is also elliptic.

We expand the denominator of \eref{eSsc} in powers of $S^z$,
\begin{equation}
\fl
S_{\u}(\lambda)=\sum_{r=0}^{\infty}\frac{(E^h_{\dd}-E^h_{\u})^r}
{\Bigl((S+1)E^h_{\u}+SE^h_{\dd}\Bigr)^{r+1}}\,
\frac{\partial^r}{\partial\lambda^r}
\la \e^{\lambda S^z}\Bigl((S+1+S^z)n^-_{\dd}+S^-\s^+\Bigr)\ra.
\label{eSexpand}
\end{equation}
Owing to the presence of the factor $E^h_{\dd}-E^h_{\u}$, which is zero in
the $h=0$ paramagnetic state, only the $r=0$ and $r=1$ terms contribute to
$\delta S_{\u}(\lambda)$. From \eref{eSexpand} we find
\begin{eqnarray}\nonumber
\fl\delta S_{\u}(\lambda)=
\frac{\delta E^h_{\dd}(\partial/\partial\lambda-S)
-\delta E^h_{\u}(\partial/\partial\lambda+S+1)}
{(2S+1)^2 E^2}
\la\e^{\lambda S^z}\left((S+1+S^z)n^-_{\dd}+S^-\s^+\right)\ra\\
+\frac{\delta\!
\la\e^{\lambda S^z}\left((S+1+S^z)n^-_{\dd}+S^-\s^+\right)\ra}
{(2S+1)E}\label{edeltaS}
\end{eqnarray}
where the paramagnetic state value, which may easily be evaluated,
is used for the first expectation. Now it may be shown that $\delta\mu=0$
so that $\delta{\rm I}[A]={\rm I}[\delta A]$ for any $A$. We
calculate $\delta{\rm I}[\partial/\partial\lambda
(S_{\u}(\lambda)+S_{\dd}(\lambda))]_{\lambda=0}$ from \eref{edeltaS}, and
using \eref{eid1} and \eref{eid2} and the sum rule \eref{esumruleS} find
\begin{eqnarray}
\fl\nonumber
\delta\la S^z (n_{\u}\uu+n_{\dd}\uu)\ra=
2S\,\delta\la \s^z\ra=\frac{n}{S+1-n/2}
\Bigl((S+1)\delta\la L^z\ra-(2S+1)\delta\la \s^z\ra\Bigr)\\
+\frac{2S\left(S+1-n/2\right)}{3(2S+1)^2}{\rm I}
\left[\frac{\delta E^h_{\dd}-2(S+1)
\,\delta E^h_{\u}}{E^2}\right].
\label{edSz}
\end{eqnarray}
Since $E^h_{\s}(\en)= \en+h/2-W^2/4\,G_{\s}(\en+h\delta_{\s\dd})$ we have
\begin{equation}\label{edeltaE}
\delta E^h_{\s}=h/2-W^2/4\,\left(\delta G_{\s}+h\delta_{\s\dd}\,G'\right)
\end{equation}
where $G'(\en)=\d G(\en)/\d\en$. By setting $\lambda=0$ in \eref{edeltaS}
and using \eref{eid1}, \eref{eid2} and \eref{edeltaE} we obtain
\begin{equation}
\delta G_{\s}=\s\,\frac{\delta\la L^z\ra E+
h\,(\bW/W)^2\Bigl(SW^2 G'/6-(S+1/2)\Bigr)}
{(2S+1)E^2-(2/3\,S+1)\bW^2/4}. \label{edG}
\end{equation}
Note that since $\delta G_{\s}\propto\s$ spectral weight is transferred
between the different spin bands at constant energy, so $\delta\mu=0$ as
mentioned above. \Eref{edG} illustrates how $\delta\las L^z\ras$, and
hence $\delta\las S^z\ras$, is determined indirectly in terms of the
Green functions rather than by a direct sum rule of the type
\eref{esumruleS}--\eref{esumruleT}.

We use \eref{edeltaE} and \eref{edG} to eliminate the $\delta E^h_{\s}$s
from \eref{edSz} in terms of $h$ and $\delta\las L^z\ras$. By applying I to
\eref{edG} we obtain an expression for $\delta\las\s^z\ras$. We then
eliminate $\delta\las\s^z\ras$ between this equation and \eref{edSz} and
solve the resulting equation for $\chi=(\gmub)^2\delta\las L^z\ras/h$,
\numparts
\begin{eqnarray}
\label{echi}
\chi=-\frac{(\gmub)^2}{2n}\Bigl(4/3\,S(S+1)+(4/3\,S+1)n\Bigr)
\frac{{\rm I}[G']}{Q-1}\\
Q={\rm I}\left[\frac{E}{E^2-\nu^2\,\bW^2/4}\right]\label{eQ}
\end{eqnarray}
\endnumparts
where $\nu^2=(2/3\,S+1)/(2S+1)$. Note that
for $S=1/2$ this expression for $\chi$ does reduce to the one given (for
the $S=1/2$ case) in \cite{rEdwardsGreenKuboConference}.
We can simplify $Q$ by changing variables to $z=W^2 G(\en)/(2\bW)$
so that $\d\en=2\bW/W^2\,G'(\en)^{-1}\d z=
\bW/2\,(1-(2\bW/W^2)^2 G(\en)^{-2})\d z=\bW/2\,(1-z^{-2})\d z$, and from
the functional form \eref{eGpara} of $G$ it may be seen that the contour
$\gamma$ for $\en$ becomes $-\gamma'$, the clockwise unit circle, for
$z$. Hence
\begin{equation}
Q=\oint_{\gamma'}\uu\frac{\d z}{2\pi i}
\bar{f}\left(\frac{z+z^{-1}}{2}-\bmu\right) \label{eQint}
\frac{z-z^{-1}}{z^2-\nu^2} 
\end{equation}
where $\bmu=\mu/\bW$ and $\bar{f}(\en)=f(\bW\en)$.
The same change of variables can be used to show that
\begin{equation}
{\rm I}[G']=-\frac{2\bW}{W^2}\oint_{\gamma'}\uu\frac{\d z}{2\pi i}
\bar{f}\left(\frac{z+z^{-1}}{2}-\bmu\right). \label{echitop}
\end{equation}

\begin{figure}[ht]
\begin{center}
\leavevmode
\hbox{%
\epsfxsize=0.45\textwidth
\epsffile{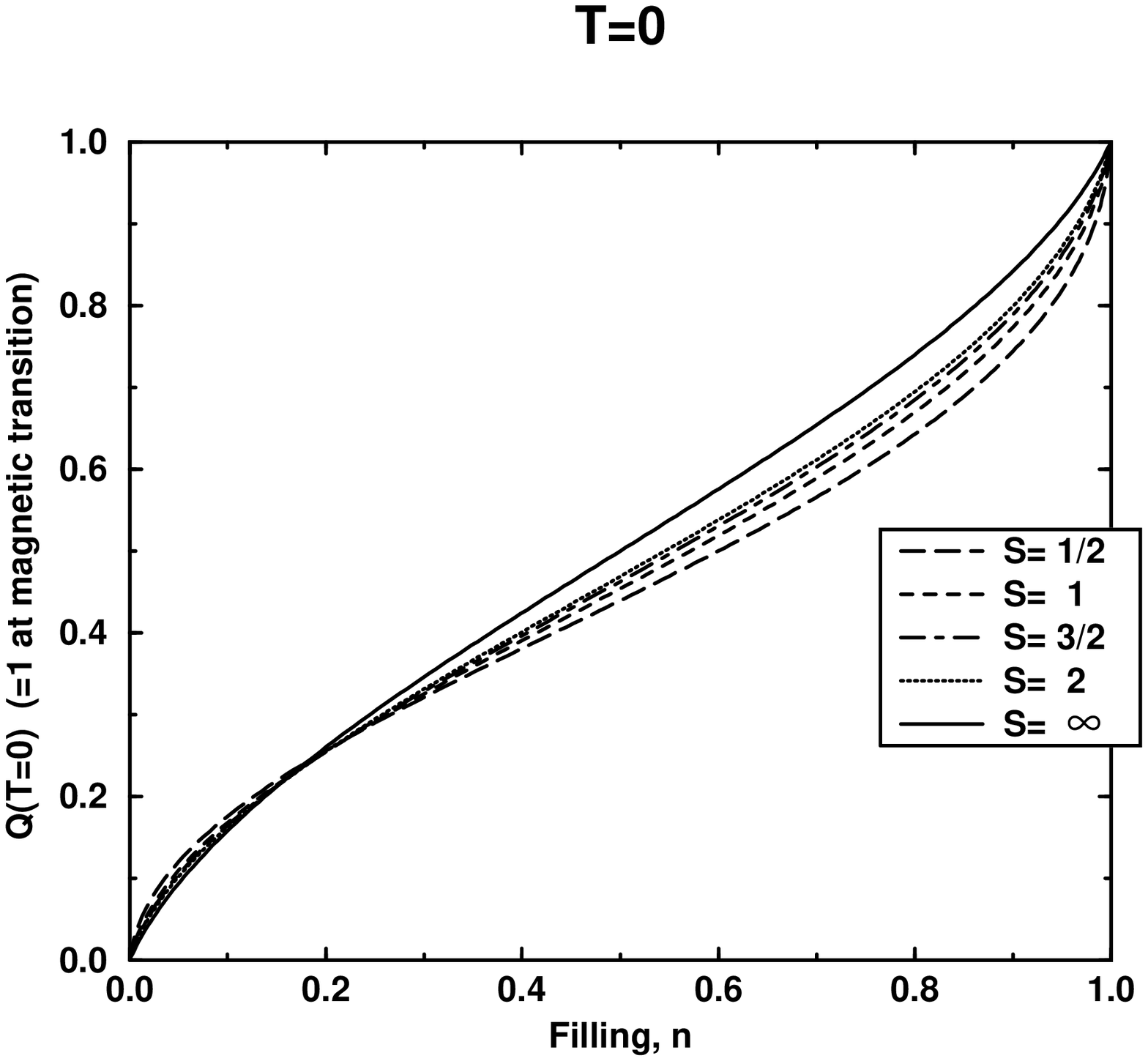}
\epsfxsize=0.45\textwidth
\epsffile{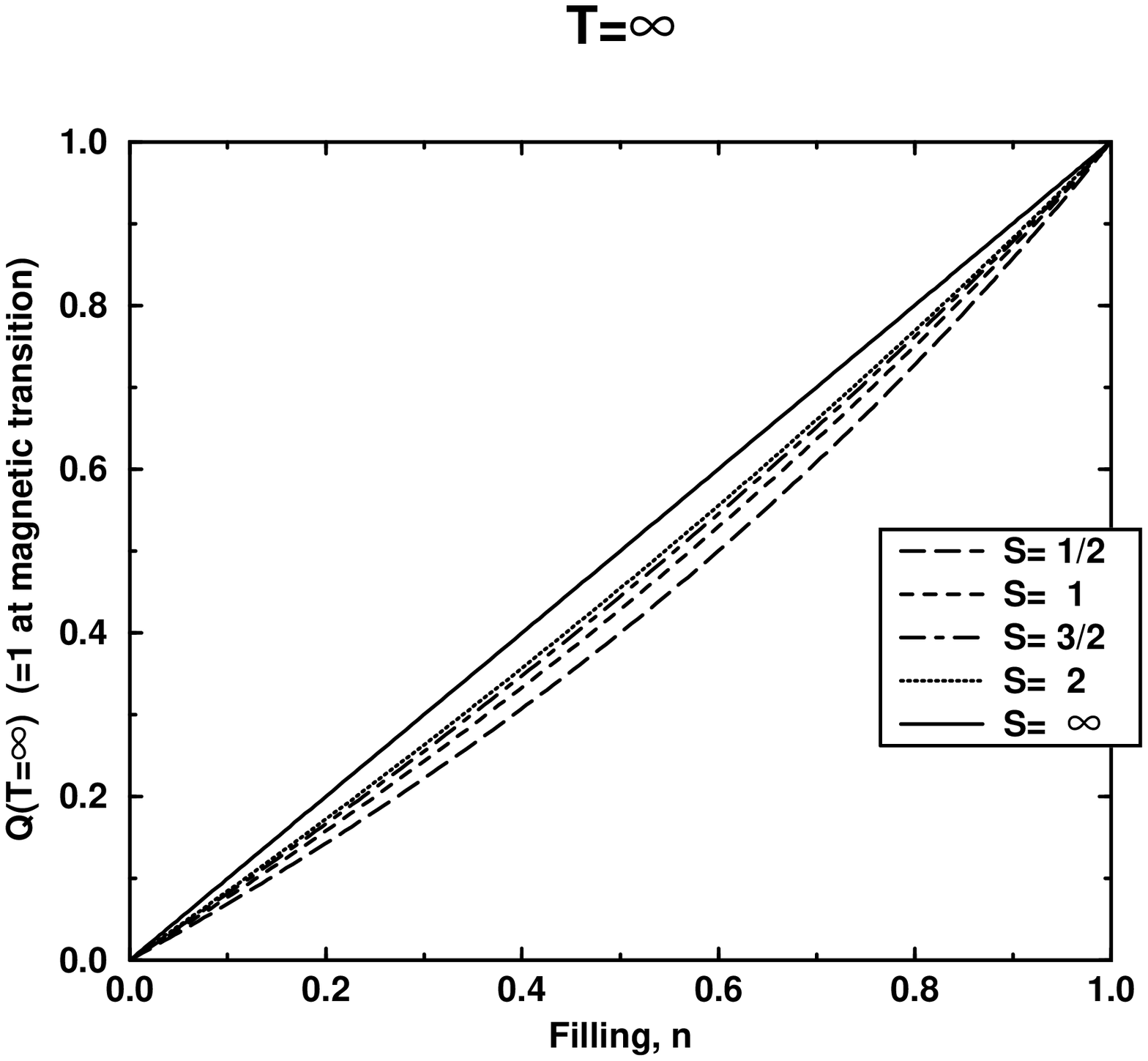}
}
\end{center}
\caption{The $Q$ function versus filling $n$ at temperature $T=0$
and $T=\infty$.\label{fQ}}
\end{figure}

Now a second-order transition to ferromagnetism corresponds to a
divergence in
$\chi$, i.e.\ $Q=1$. From \eref{edG} with $h=0$ and \eref{eQ} it may be
seen, using the sum rule, that $\delta\las L^z\ras Q/(2S+1)={\rm I}
[\delta G_{\u}]=\delta\las\s^z\ras$ (for $h=0$). The equation $Q=1$ for a
zero field magnetic transition is therefore equivalent to the consistency
condition $2\delta\las\s^z\ras=\delta\las S^z+\s^z\ras/(S+1/2)$, which
certainly holds
at $n=1$ but not at $n=0$. Integral \eref{eQint}
can be evaluated analytically in the limits of zero and infinite
temperature where the Fermi function is of a simple form and the results
are plotted in \fref{fQ}. It is clear that within the CPA there is no
magnetic transition for $0<n<1$, as is the case for the CPA for the Hubbard
model \cite{rFukuyama}. This appears to be a considerable drawback of our
approximation given that the ferromagnetic-paramagnetic phase transition
is a major reason for interest in the double exchange model. We will
propose a method for circumventing this problem in \sref{sDMFTTc}.

\begin{figure}[ht]
\begin{center}
\leavevmode
\hbox{%
\epsfxsize=0.8\textwidth
\epsffile{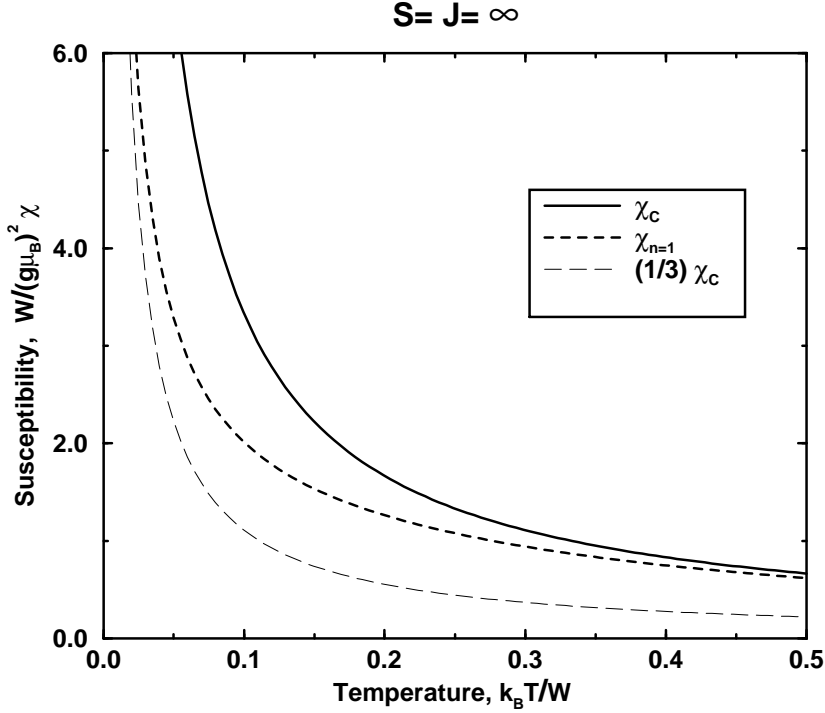}
}
\end{center}
\caption{The $S=J=\infty$ magnetic susceptibility $\chi$ for $n=1$
compared with its low temperature asymptote and the Curie law.
\label{fsuscomp1}}
\end{figure}

\begin{figure}[ht]
\begin{center}
\leavevmode
\hbox{%
\epsfxsize=0.8\textwidth
\epsffile{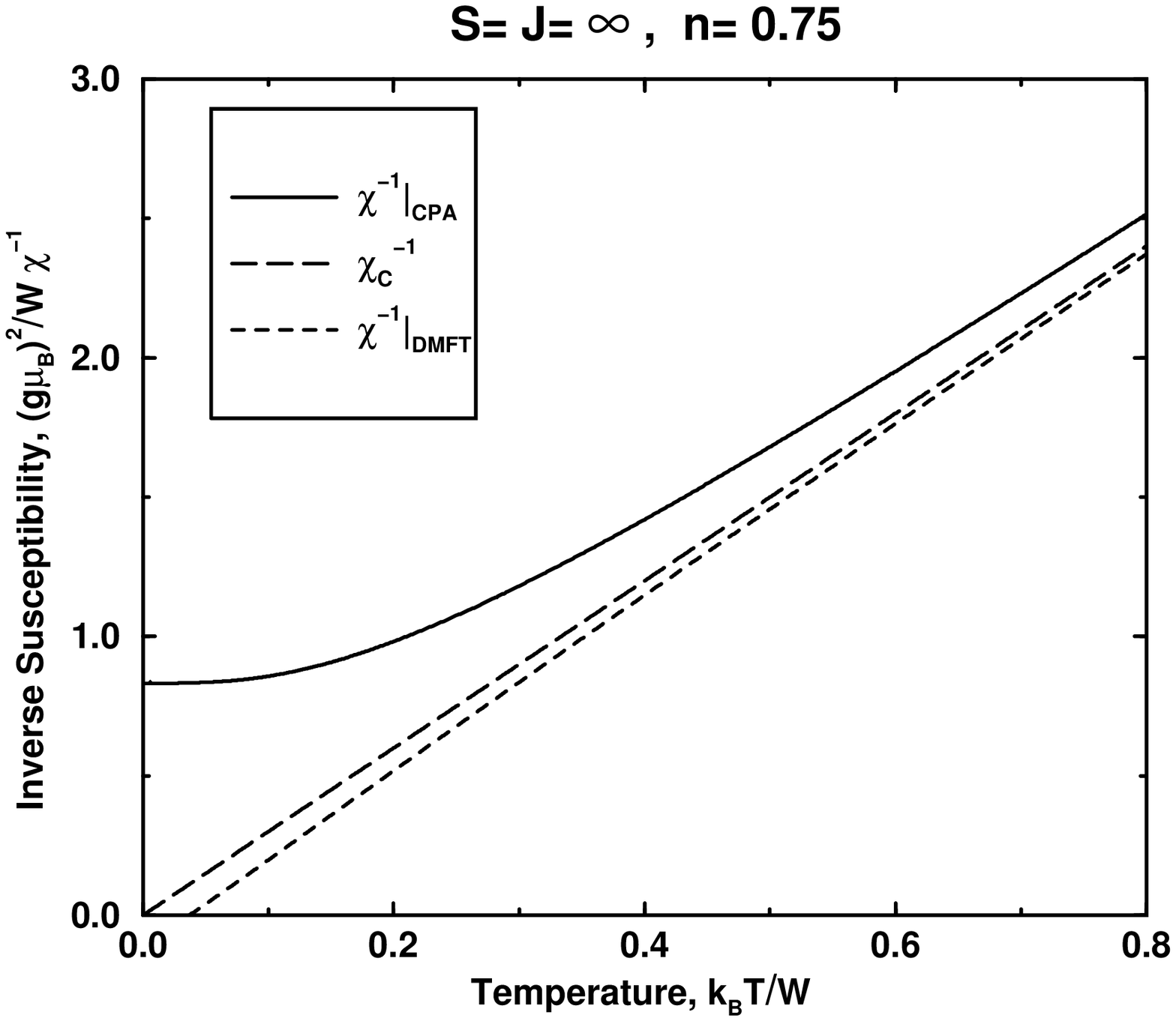}
}
\end{center}
\caption{The $S=J=\infty$ inverse magnetic susceptibility $\chi^{-1}$
for $n=0.75$ compared with the Curie law and DMFT values.\label{fsuscomp2}}
\end{figure}

We now consider the behaviour of $\chi$ at $n=0,1$. In these
cases electron hopping does not occur
and the system consists of a
lattice of free local moments of magnitude $S'=S$ and $S+1/2$
respectively, so we expect $\chi$ to take the Curie law form,
$\chi_{\rm C}=(\gmub)^2\beta S'(S'+1)/3$ where $\beta=(\kB T)^{-1}$ is the
inverse temperature. We calculate $\chi$ in these
cases by expanding the integrals in \eref{eQint} and \eref{echitop} in
powers of $n$ or $1-n$. At $n=0$ we find $\chi=\chi_{\rm C}$ with $S'=S$
as expected, but at $n=1$ we find $\chi=\chi_{\rm C}\phi$ (with $S'=S+1/2$)
where
\begin{equation}
\phi=\frac{\oint_{\gamma'}\d z\exp(\bar\beta(z+z^{-1})/2)}
{\oint_{\gamma'}\d z\exp(\bar{\beta}(z+z^{-1})/2)\,
(z-z^{-1})/(z^2-\nu^2)}
\end{equation}
and $\bar\beta=\bW\beta$. Now $\phi\rightarrow 1$ as $\beta\rightarrow 0$
so the Curie law is obtained at high temperature, but as
$\beta\rightarrow\infty$ we find
$\phi\rightarrow 8 S^2/(3(2S+1)(4S+3))\ne 1$ so Curie
law behaviour does not extend over the whole temperature range. 
Note that $\phi(\beta=\infty)\ne 1$ even at
$S=\infty$ where, as will be shown in the next section,
our Green function equations reduce to DMFT. The reasons for this
unphysical behaviour, and a way to avoid it,
are discussed in \sref{sDMFTTc}.
In \fref{fsuscomp1} below we compare $\chi$ at $n=1$ and
$S=\infty$ with $\chi_{\rm C}$ and the low temperature asymptote
$\chi_{\rm C}/3$. In \fref{fsuscomp2} we plot $\chi^{-1}$ at $S=\infty$ and
$n=0.75$, comparing it with the Curie law and DMFT values, the DMFT plot
only being displayed for the paramagnetic phase.

\section{Comparison with dynamical mean field theory at
\mbox{\boldmath$S=\infty$}}
\label{sDMFT}

In this section we will obtain equations for the $S^{\alpha}_{\s}$ and
$T^{\alpha}_{\s}$ Green
functions within dynamical mean field theory (DMFT) at $S=\infty$ and
compare them with equations \eref{eSSinf} and \eref{eTSinf} of the CPA.
The DMFT single-site effective action $\tilde S$ of the DE model is
\cite{rFurukawa}
\begin{eqnarray}
\fl\nonumber
\tilde{S}= \int_{0}^{\beta}\int_{0}^{\beta}\d\tau\d\tau'\sum_{\s}
c^{\dag}_{\s}(\tau)E_{\s}(\tau-\tau')\/c_{\s}(\tau')-\frac{J}{2}
\int_{0}^{\beta}\d\tau\Bigl[S^z(\tau)\left[n_{\u}(\tau)-n_{\dd}(\tau)
\right]\Bigr.\\
\Bigl.+\,S^+(\tau)\s^-(\tau)+S^-(\tau)\s^+(\tau)\Bigr]
-h\int_0^{\beta}\d\tau\Bigl[S^z(\tau)+\s^z(\tau)\Bigr]
\end{eqnarray} 
where $E_{\s}$ describes the self-consistently determined coupling with the
conduction electron bath. Here $c^{\dag}_{\s}(\tau)$ and 
$c_{\s}\uu(\tau)$ are Grassmann variables \cite{rNegele}.
Now DMFT for the DE model is exactly solvable in
the classical spin limit $S\rightarrow\infty$ where $\S(\tau)$ becomes
$\tau$-independent (we scale $J$ and $h$ as $1/S$), since for $\S$
constant $\tilde S$ is diagonal in the Matsubara frequency representation:
\numparts
\begin{eqnarray}
\fl
\tilde{S} =-\sum_n
\left(\begin{array}{lr}
c^{\dag}_{n\u}\mbox{ } c^{\dag}_{n\dd}
\end{array}\right)
\left[\begin{array}{cc}
E_{n\u}+J/2\,S^z & J/2\,S^-\\
J/2\,S^+ & E_{n\dd}-J/2\,S^z
\end{array}\right]
\left(\begin{array}{c}
c_{n\u}\\
c_{n\dd}
\end{array}\right)-\beta hS^z\\
\hskip -0.25in
\mbox{ }=\sum_n\left(\begin{array}{lr}
c^{\dag}_{n\u}\mbox{ } c^{\dag}_{n\dd}
\end{array}\right)A_n\left(\begin{array}{c}
c_{n\u}\\
c_{n\dd}
\end{array}\right)-\beta h S^z= \sum_n\tilde{S}_n-\beta hS^z
\end{eqnarray}
\endnumparts
defining the matrix $A_n$ and the components $\tilde{S}_n$ of 
$\tilde{S}$. Here $E_{n\s}=-\int_0^{\beta}\d\tau
\exp(i\omega_n\tau)E_{\s}(\tau)$ and
$\omega_n$ is a fermionic Matsubara frequency.
Note that as shown in
\cite{rDMFT} the self-consistency condition for $E_{n\s}$ can be written as
$E_{n\s}=\Sigma(i\omega_n)+G(i\omega_n)^{-1}$, so $E_{n\s}$ is just our
quantity $E_{\s}(i\omega_n)$.

In terms of $\tilde S$ the partition function $Z$ is given
by
\begin{equation}
\fl Z=\int\d^2 S\int\Bigl(\prod_{n\s}\d c^{\dag}_{n\s}\d c\uu_{n\s}\Bigr)
\e^{-{\tilde S}}=
\int\d^2 S\,\e^{\beta hS^z}\prod_{n}\int
\Bigl(\prod_{\s}\d c^{\dag}_{n\s}\d c\uu_{n\s}\Bigr)\e^{-{\tilde S}_n}
\end{equation}
where $\int\d^2 S$ is the integral over the local spin direction.
All site-diagonal correlation functions can be
calculated explicitly in terms of the $E_{n\s}$s, for example the one
electron Green function is given by
\begin{equation}
\fl
\lla c_{i\u}\uu\,;c^{\dag}_{i\u}\rra_{i\omega_n}=
-\la c_{n\u}\uu c^{\dag}_{n\u}\ra=-\frac{1}{Z}\int\d^2 S
\Bigl(\prod_{m\s}\d c^{\dag}_{m\s}\d c\uu_{m\s}\Bigr)\e^{-{\tilde S}}
c_{n\u}\uu c^{\dag}_{n\u}.
\end{equation}

It is convenient to work with the generating function
\numparts
\begin{eqnarray}
Z_n &=\int\left(\prod_{\s}\uu\d c^{\dag}_{n\s}\d c_{n\s}\uu\right)
\exp\left(-\tilde{S}_n+\sum_{\s}\left(\eta^{\dag}_{n\s}c_{n\s}\uu+
c_{n\s}^{\dag}\eta_{n\s}\uu\right)\right)\\
\mbox{ }&=\det(A_n)\exp\left[
\left(\begin{array}{lr}
\eta^{\dag}_{n\u}\mbox{ } \eta^{\dag}_{n\dd}
\end{array}\right)A_n^{-1}
\left(\begin{array}{c}
\eta_{n\u}\\
\eta_{n\dd}
\end{array}\right)\right].
\end{eqnarray}
\endnumparts
In terms of $Z_n$ the partition function $Z=\int\d^2 S\,\exp(\beta h S^z)
\left(\prod_n Z_n\right)_{\eta'{\rm s}=0}=
\int\d^2 S\,\exp(\beta h S^z)\prod_n\det(A_n)$ and the local spin
probability distribution function
$P(\S)=Z^{-1}\exp(\beta h S^z)\left(\prod_n Z_n\right)_{\eta'{\rm s}=0}$.
Note that $\las U(\S)\ras=\int\d^2 S\,P(\S)U(\S)$ for any
$U$.

Explicitly, our full Green functions are given in $S=\infty$ DMFT by
\numparts
\begin{eqnarray} 
S_{\u}(\lambda,i\omega_n)=\frac{1}{Z}\int\d^2 S\,\e^{(\lambda+\beta h)S^z}
\left(\frac{\d^2}{\d\eta^{\dag}_{n\u}\d\eta_{n\u}\uu}
\prod_{n'}Z_{n'}\right)_{\eta'{\rm s}=0}\\
T_{\u}(\lambda,i\omega_n)=\frac{1}{Z}\int\d^2 S\,
\e^{(\lambda+\beta h)S^z}S^-
\left(\frac{\d^2}{\d\eta^{\dag}_{n\dd}\d\eta_{n\u}\uu}
\prod_{n'}Z_{n'}\right)_{\eta'{\rm s}=0},
\end{eqnarray}
\endnumparts
and these expressions are easily evaluated to give
\numparts
\begin{eqnarray}
\fl S_{\u}(\lambda,i\omega_n)=\la\e^{\lambda S^z}
\frac{E_{n\dd}-J/2\,S^z}
{\left(E_{n\u}+J/2\,S^z\right)\left(E_{n\dd}-J/2\,S^z\right)
-J^2/4\,\left(1-\left(S^z\right)^2\right)}\ra \label{eDMFTS}\\
\fl T_{\u}(\lambda,i\omega_n)=-\frac{J}{2}
\la\e^{\lambda S^z}\frac{1-\left(S^z\right)^2}
{\left(E_{n\u}+J/2\,S^z\right)\left(E_{n\dd}-J/2\,S^z\right)
-J^2/4\,\left(1-\left(S^z\right)^2\right)}\ra. \label{eDMFTT}
\end{eqnarray}
\endnumparts
Summing the $S=\infty$ CPA expressions \eref{eSSinf} and \eref{eTSinf} over
$\alpha$ we obtain the analytic continuations (from the Matsubara
frequencies) of \eref{eDMFTS} and
\eref{eDMFTT} respectively, i.e.\ in the classical spin limit our CPA
agrees with DMFT for $S_{\s}(\lambda)$ and $T_{\s}(\lambda)$.
This is important since DMFT
is known to be exact for dimension $D=\infty$ \cite{rDMFT} and is perhaps
the most natural local approximation for $D$ finite.

Similarly we can calculate the $\alpha=\pm$ components of these Green
functions within DMFT as
\numparts
\begin{eqnarray}
\fl S^+_{\u}(\lambda,i\omega_n)=\frac{1}{\beta}\sum_{n'}
\la\e^{\lambda S^z}\frac{\left(E_{n\dd}-J/2\,S^z\right)
\left(E_{n'\u}+J/2\,S^z\right)-J^2/4\,\left(1-\left(S^z\right)^2\right)}
{\det(A_n)\det(A_{n'})}\ra \label{eDMFTS+}\\
\fl T^+_{\u}(\lambda,i\omega_n)=\frac{J}{2}\frac{1}{\beta}\sum_{n'}
\la\e^{\lambda S^z}\left(1-\left(S^z\right)^2\right)
\frac{E_{n\dd}-E_{n'\dd}}{\det(A_n)\det(A_{n'})}\ra \label{eDMFTT+}
\end{eqnarray}
\endnumparts
and $S^-_{\u}=S_{\u}\uu-S^+_{\u}$ etc. Now for a function
$g(z)$ analytic off the real axis $\beta^{-1}\sum_n g(i\omega_n)=I[g]$, so
applying this to the $n'$ summations in \eref{eDMFTS+} we find
\begin{eqnarray}
\fl\nonumber S^+_{\u}(\lambda,i\omega_n)=\\
\hskip -0.5in\la\e^{\lambda S^z}
\frac{(E_{n\dd}-JS^z/2)(I_{1\u}(S^z)+JS^z I_2(S^z)/2)-
J^2/4\Bigl(1-(S^z)^2\Bigr)I_2(S^z)}
{\left(E_{n\u}+J/2\,S^z\right)\label{eSproc}
\left(E_{n\dd}-J/2\,S^z\right)-J^2/4\,\left(1-\left(S^z\right)^2\right)}\ra
\end{eqnarray}
where
\numparts
\begin{eqnarray}
\fl
I_{1\u}(S^z)=
I\left[\frac{E_{\u}}{\left(E_{\u}+J/2\,S^z\right)
\left(E_{\dd}-J/2\,S^z\right)-J^2/4\,\left(1-\left(S^z\right)^2\right)}
\right]\\
\fl
I_2(S^z)=
I\left[\frac{1}{\left(E_{\u}+J/2\,S^z\right)
\left(E_{\dd}-J/2\,S^z\right)-J^2/4\,\left(1-\left(S^z\right)^2\right)}
\right].
\end{eqnarray}
\endnumparts
It may be shown using the sum rules \eref{esumruleS} and \eref{esumruleT}
that $\las (S^z)^m(I_{1\u}(S^z)+JS^zI_2(S^z)/2)\ras=
\las(S^z)^m n_{\dd}\ras$
and $J/2\,\las (S^z)^m(1-(S^z)^2)I_2(S^z)\ras=-\las(S^z)^mS^-\s^+\ras$ for
any $m$, so in \eref{eSproc}
we can replace $I_{1\u}(S^z)+JS^zI_2(S^z)/2$ with
$n_{\dd}$ and $J/2\,(1-(S^z)^2)I_2(S^z)$ with $-S^-\s^+$. A similar
procedure can be carried out for \eref{eDMFTT+}, and we obtain
\numparts
\begin{eqnarray}
\fl S^+_{\u}(\lambda,i\omega_n)=
\la\e^{\lambda S^z}\frac{\left(E_{n\dd}-J/2 S^z\right)n_{\dd}
+J/2\,S^-\s^+}{\left(E_{n\u}+J/2\,S^z\right)\left(E_{n\dd}-J/2\,S^z\right)
-J^2/4\left(1-\left(S^z\right)^2\right)}\ra\\
\fl T^+_{\u}(\lambda,i\omega_n)= \la\e^{\lambda S^z}
\frac{J/2\left(\left(S^z\right)^2-1\right)n_{\u}-
\left(E_{n\dd}-J/2\, S^z\right)S^-\s^+}
{\left(E_{n\u}+J/2\,S^z\right)\left(E_{n\dd}-J/2\,S^z\right)
-J^2/4\left(1-\left(S^z\right)^2\right)}
\ra.
\end{eqnarray}
\endnumparts
It may be seen that the DMFT and CPA expressions for $S^+_{\s}$ agree,
but $T^+_{\u}(\lambda)\vert_{\rm DMFT}=T^+_{\dd}(\lambda)\vert_{\rm CPA}$.
This discrepancy, due to failure of the decoupling
approximations made in the equation of motion for $T^{\alpha}_{\s}$
\cite{rEdwardsGreenKubo}, vanishes in the important limit
$J\rightarrow\infty$ of strong coupling where all $\alpha=+$ Green
functions are zero.

\section{The Curie temperature}\label{sDMFTTc}

Furukawa \cite{rFurukawa} finds that the $S=J=D=\infty$ DE model exhibits
a transition to ferromagnetism, and in \sref{sDMFT} it was shown that our
CPA gives exact expressions for the Green function equations in this
limit. However, in \sref{sCPAsus} we showed that there is no magnetic
transition in our CPA (\fref{fQ}) and found unphysical behaviour at $n=1$
(\fref{fsuscomp1}). In this section we resolve this apparent discrepancy
and postulate a
modification of our CPA that restores magnetic behaviour. We again work in
the elliptic DOS case where $E_{n\s}=i\omega_n-W^2/4\,G_{\s}(i\omega_n)$.

We first derive the $S=J=\infty$ DMFT static susceptibility $\chi$.
In this case
\begin{equation}
P\left(\S\right)=\frac{\exp\left(\beta h S^z\right)\prod_n\left[
\left(1+S^z\right)E_{n\u}+\left(1-S^z\right)E_{n\dd}\right]}
{\int\d^2 S \exp\left(\beta h S^z\right)\prod_n\left[
\left(1+S^z\right)E_{n\u}+\left(1-S^z\right)E_{n\dd}\right]},\label{ePDMFT}
\end{equation}
as was shown in the previous section.
The first order deviation $\delta P(\S)$ of $P(\S)$ from its
zero-field paramagnetic state value ($1/4\pi$) is given by
\begin{equation}
\delta P(\S)=\left(\beta h+
\sum_n\frac{\delta E_n}{E_n}\right)\frac{S^z}{4\pi} \label{edPDMFT}
\end{equation}
where $E_n$ is the zero-field paramagnetic state value and $\delta E_n$
is the first order deviation of $E_{n\u}$. From \eref{edG} we have in the
present case, where $\bW=W/\sqrt{2}$,
\begin{equation}
\delta G_{n\s}= \frac{\s E_n\delta\las S^z\ras}{2E_n^2-W^2/12}.
\label{edGDMFT}
\end{equation}
Substituting \eref{edGDMFT} into \eref{edPDMFT} using $\delta E_{n\s}=
-W^2/4\,\delta G_{n\s}$, then multiplying by $S^z$ and integrating over the
local spin orientation we obtain
\begin{equation}
\chi=\frac{\gmub\delta\las S^z\ras}{B}=
\frac{(\gmub)^2\beta/3}{1+W^2/12\,\sum_n\left(2E_n^2-W^2/12\right)^{-1}}
\end{equation}
upon rearrangement.
This gives the correct Curie law $\chi=1/3\,(\gmub)^2\beta$ at $n=0$ and
$n=1$ and a transition to ferromagnetism for all $n$ at a temperature
$k_{\rm B}T=-W^2/12\,I[(2E^2-W^2/12)^{-1}]$ where
$E(\en)=\en-W^2/4\,G(\en)$ \cite{rFurukawa}.

We now consider where the CPA calculation of $\chi$ has gone wrong
and how to improve it.
The CPA equations for the Green functions are exact in the present
case ($S=J=\infty$), but to solve them we need expressions for the
expectations $\las\exp(\lambda S^z) n_{\s}\ras$,
$\las\exp(\lambda S^z) S^{-\s}\s^{\s}\ras$, and
$\las\exp(\lambda S^z)\ras$, as mentioned in \sref{sCPAsoln}. Since the
first two of these are obtained using the I sum rule, a procedure that is
exact, the problem must lie with the determination of
$\las\exp(\lambda S^z)\ras$. Note that the way that we have calculated
$\las\exp(\lambda S^z)\ras$ only works for $S$ finite, and to calculate it
at $S=\infty$ we have worked for finite $S$ and taken the limit at the
end.

Now knowledge of $\las\exp(\lambda S^z)\ras$ is equivalent to knowledge of
$P(\S)$, so a possible way of improving the CPA expression for $\chi$ 
is to abandon the above self-consistent determination of
$\las\exp(\lambda S^z)\ras$ and instead to use some expression for $P(\S)$
that reduces to the $S=\infty$ DMFT result \eref{ePDMFT} in the classical
spin limit. We have so far been unable to derive such an expression,
so we instead postulate a natural extrapolation of \eref{ePDMFT} to finite
$S$, justifying our formula by the resulting behaviour of $\chi$.
This procedure will at least force the CPA for $\chi$ to become
exact in the $S\rightarrow\infty$ limit, and the $S=\infty$
magnetic transition is likely to persist to finite $S$.  

From the $S\rightarrow\infty$ limit of \eref{eSsc} and \eref{eTsc} it may
be seen that the quantity in square brackets in \eref{ePDMFT} is related
to the denominators of the Green function equations (for $S=\infty$).
A natural extension of \eref{ePDMFT} is thus
\begin{equation}
\fl
P\left(\S\right)=
\frac{\exp\left(\beta h\eta S^z\right)
\prod_n\left[\left(1/2+S+S^z\right)E_{n\u}+
\left(1/2+S-S^z\right)E_{n\dd}\right]}
{2\pi\sum_{S^z}\exp\left(\beta h\eta S^z\right)
\prod_n\left[\left(1/2+S+S^z\right)E_{n\u}+
\left(1/2+S-S^z\right)E_{n\dd}\right]} \label{eguessP}
\end{equation}
where $\eta$ is chosen so as to optimise the behaviour of $\chi$.
The explicit energy shifts associated with the field $h$ in $E_{\u}$ and
$E_{\dd}$ are ambiguous and have been neglected. However some effect of $h$
on the double exchange enters through the Green functions and the factor
$\eta$ allows for the conduction electron contribution to the spin in the
Zeeman energy. Then proceeding as above we can calculate 
\begin{equation}
\chi=\left(\gmub\right)^2
\frac{1/3\,\eta\beta S(S+1)-\beta sI[G'/E]+1/2\,I[G']}{1-Q/(2S+1)+
2\beta s R}\label{echiDMFT}
\end{equation}
where $s=W^2/12\,S(S+1)/(2S+1)$, $Q$ is as in \eref{eQ}, and
\begin{equation}
R=I\left[\frac{1}{(2S+1)E^2-(2/3\,S+1)\bW^2/4}\right].
\end{equation}
It is then easy to see that if we take $\eta=1+(n/2)/(S+1)$ then the
correct Curie laws $\chi=(\gmub)^2 \beta S(S+1)/3$ and
$\chi=(\gmub)^2 \beta (S+1/2)(S+3/2)/3$ are obtained at $n=0$ and $n=1$
respectively. Note that $S\eta=S+n/2+O(1/S)$, so if we regard our
extrapolation from $S=\infty$ as a kind of $1/S$ expansion this is a very
natural value--- it corresponds to the average
spin size in the system to leading order.
With this form \eref{echiDMFT} for $\chi$ we also obtain a magnetic
transition for all $S$ at a temperature determined by
\begin{equation} \label{eTc}
k_{\rm B}T=\frac{2sR}{Q/(2S+1)-1}.
\end{equation}
The Curie temperature is plotted against filling $n$ for various $S$ in
\fref{fCurie} (left figure) below. It agrees with Furukawa's result in
his case of $S=\infty$.
Clearly for finite $S$ ferromagnetism is more stable for
$n>1/2$ than for $n<1/2$, in agreement with the findings of Brunton
and Edwards \cite{rBruntonEdwards}. We have also calculated $T_{\rm C}$
via the spin-wave dispersion in the (assumed) saturated ferromagnetic
groundstate, using a method similar to that of Sakurai
\cite{rHubbardI,rSakurai} for the Hubbard model. The Curie temperature
obtained is similar in magnitude to $T_{\rm C}$ in \fref{fCurie} and
decreases with increasing $S$, as is the case here for $n$ near 1. This
work will be published elsewhere.

Brunton and Edwards
found that the stability of the spin-saturated state at $T=0$
is strongly dependent on the bare DOS used: approximating the true cubic
tight-binding DOS with the elliptic DOS qualitatively changed the form of
their spin-flip excitation gap. Accordingly we check the effect on $\Tc$
of using the true tight-binding DOS. The bare elliptic and cubic
tight-binding DOSs and the corresponding full (zero-field paramagnetic
state, $S=1$, $n=1/2$, and $J=\infty$) CPA DOSs are shown for comparison
in \fref{fbareDOSs} below. Now it is straightforward to extend the
derivation of the $\Tc$ equation to the case of a general DOS; the only
effect on \eref{eTc} is to replace $W^2/4$ with $(\bW/(WG))^2+1/G'$ inside
the I-functionals. Hence for general DOS
\begin{equation} \label{eTc2}
k_{\rm B}\Tc=\frac{2}{3}\frac{S(S+1)}{2S+1}\frac{\tilde{R}}{\tilde{Q}/
(2S+1)-1}
\end{equation}
where
\numparts
\begin{eqnarray}
\tilde{R}=I\left[\frac{(\bW/(WG))^2+1/G'}{(2S+1)E^2-(2/3\,S+1)\bW^2/W^2
((\bW/(WG))^2+1/G')}\right]\\
\tilde{Q}=I\left[\frac{E}{E^2-\nu^2\bW^2/W^2
((\bW/(WG))^2+1/G')}\right].
\end{eqnarray}
\endnumparts
Note that for the elliptic DOS case $(\bW/(WG))^2+1/G'=W^2/4$.
We plot $\Tc$
obtained from \eref{eTc2} in the most sensitive case of $S=1/2$ in
\fref{fCurie} (right figure)
for the elliptic and cubic DOSs. It may be seen that
changing the form of the bare DOS does not have a large effect on $\Tc$.
The dip in $T_{\rm C}$ near $n=0.3$ for the cubic DOS is interesting since
near this filling Brunton and Edwards \cite{rBruntonEdwards} found an
instability of the saturated ferromagnetic state for $S=1/2$.

\begin{figure}[ht]
\begin{center}
\leavevmode
\hbox{%
\epsfxsize=0.45\textwidth
\epsffile{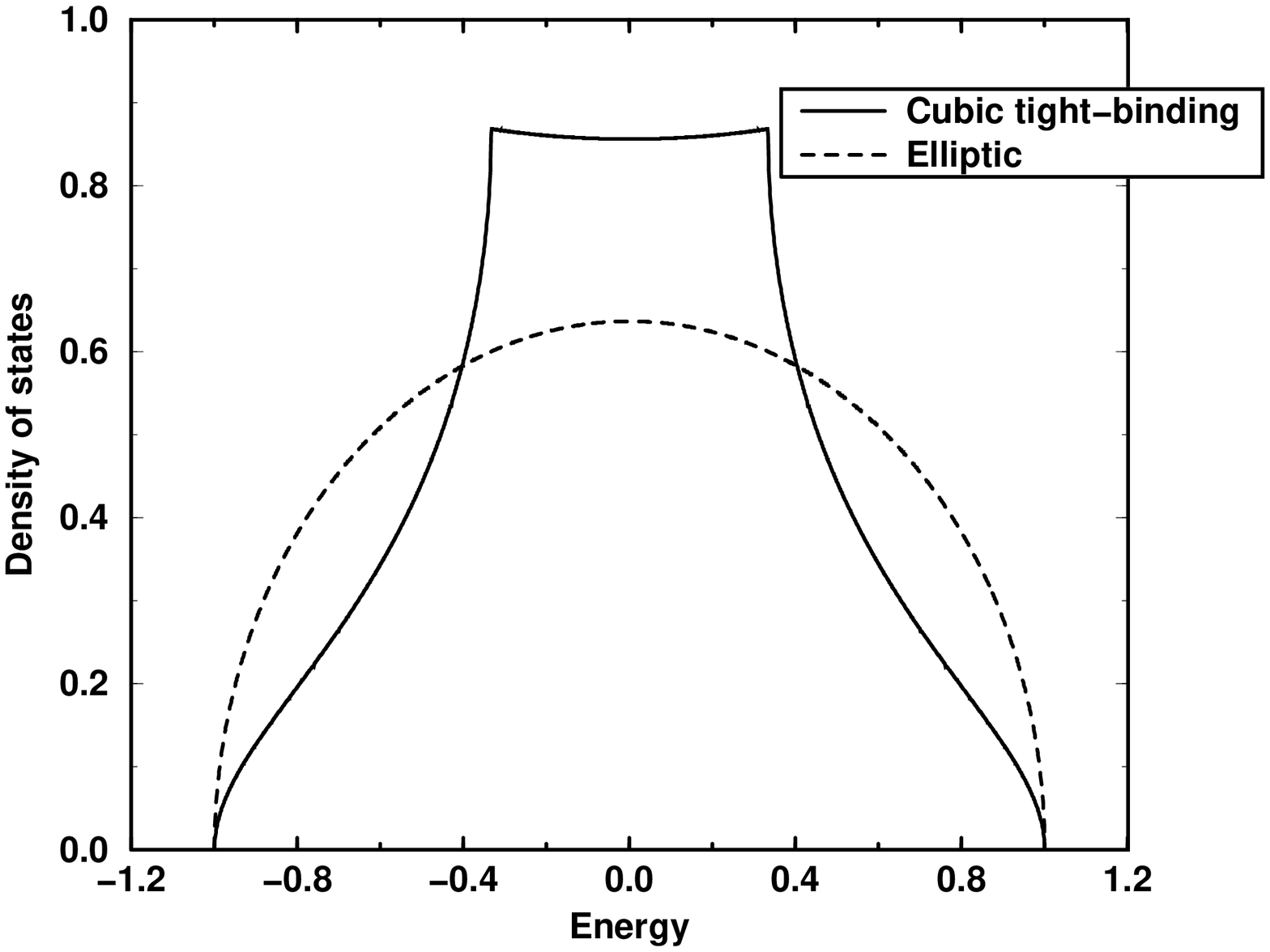}
\epsfxsize=0.45\textwidth
\epsffile{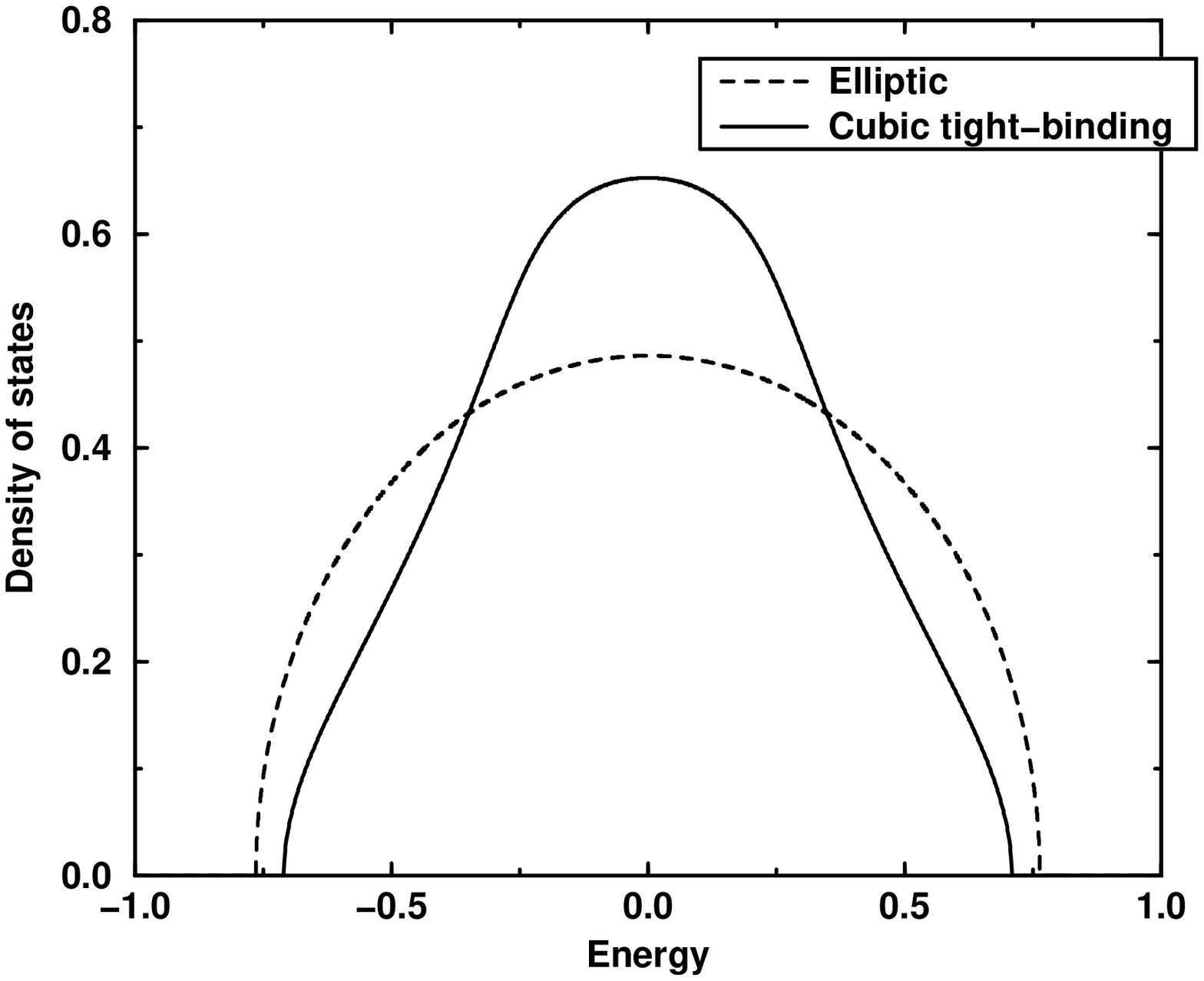}}
\end{center}
\caption{Bare (left figure) and full CPA (right figure, for $S=1$, $n=1/2$,
$J=\infty$, $W=1$ zero-field paramagnetic state) DOSs.
\label{fbareDOSs}}
\end{figure}

\begin{figure}[ht]
\begin{center}
\leavevmode
\hbox{%
\epsfxsize=0.45\textwidth
\epsffile{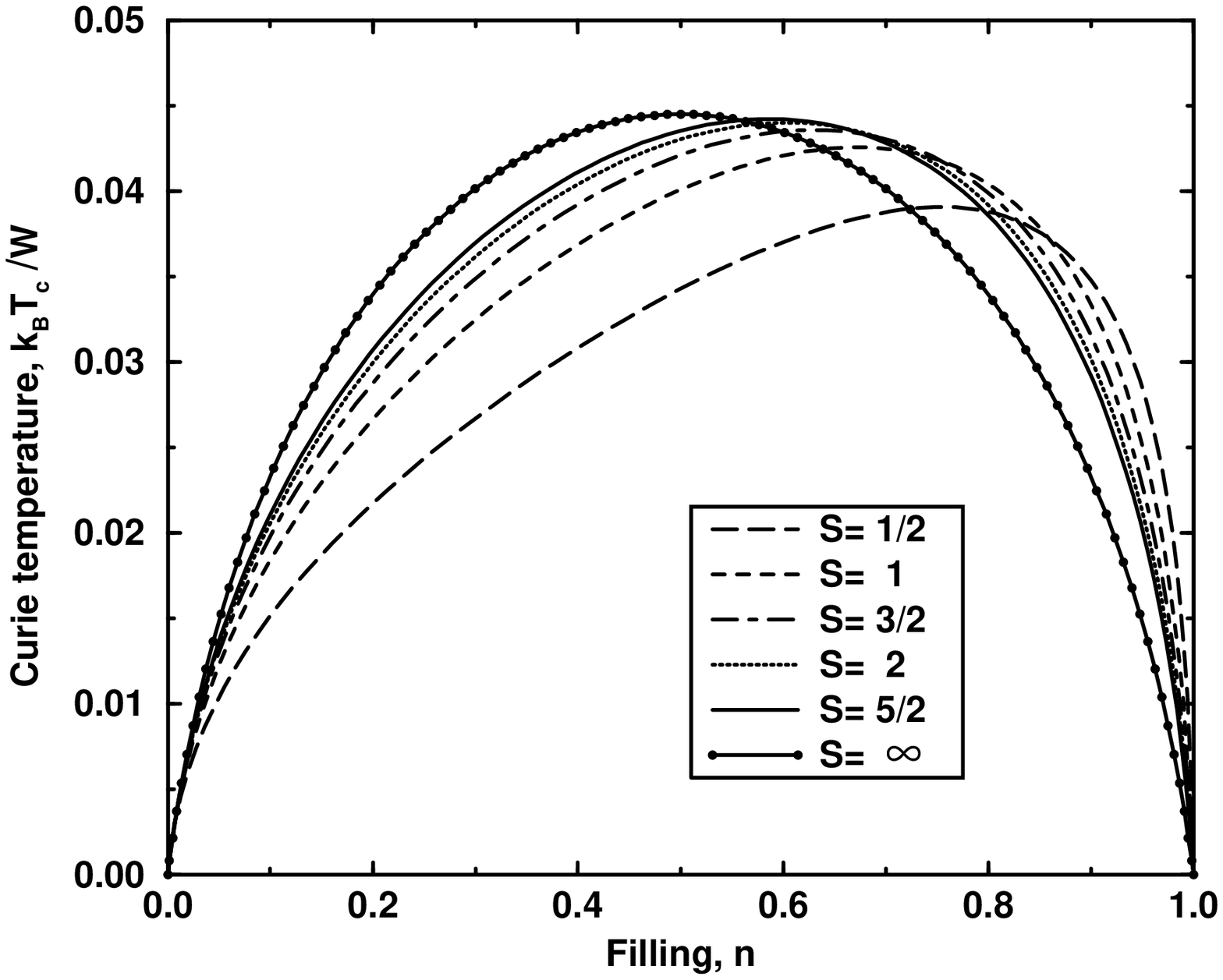}
\epsfxsize=0.45\textwidth
\epsffile{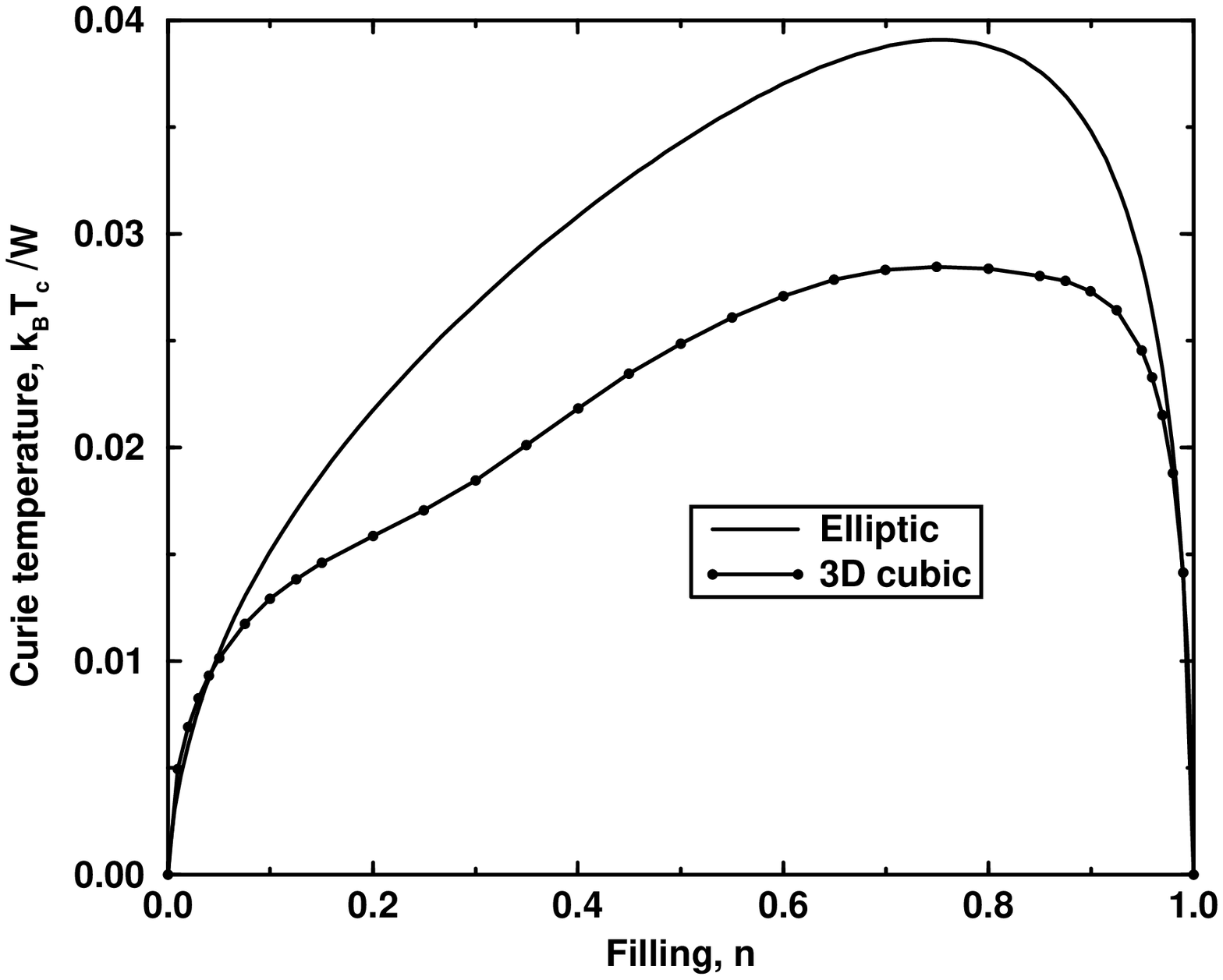}}
\end{center}
\caption{The Curie temperature $k_{\rm B}\Tc/W$ calculated using the
elliptic
bare DOS plotted against filling $n$ for various $S$ (left figure), and the
effect on $\Tc$ for $S=1/2$ of changing the bare DOS to the 3d-cubic DOS
(right figure).
\label{fCurie}}
\end{figure}

\section{Summary and outlook}\label{ssummary}

In this paper we have extended our many-body CPA treatment
\cite{rEdwardsGreenKubo} of the DE model to the case of general $S$ and
magnetisation. In our original approach we were faced with $4S+1$ algebraic
equations to solve for the Green functions in the case of non-zero
magnetisation. A correspondingly large number of correlation functions had
to be determined self-consistently. Consequently in
\cite{rEdwardsGreenKubo} we only considered $S=1/2$ for the magnetised
state and subsequently \cite{rEdwardsGreenKuboConference} calculated the
paramagnetic susceptibility in this case. The generalisation to arbitrary
$S$ in \sref{sCPAsoln} of this paper is achieved by introducing generating
Green functions involving a parameter $\lambda$. The $4S+1$ coupled
algebraic equations are then replaced by a single first order linear
differential equation in $\lambda$ whose solution yields the CPA equations
for the Green functions.  Only three correlation functions have to be
determined, as functions of $\lambda$, and two of these may be obtained
directly from the Green functions. The indirect determination of the third
$\las\exp(\lambda S^z)\ras$, from the approximate EOM for the Green
functions, is less reliable. It seems to be the origin of difficulties in
\sref{sCPAsus}, where the paramagnetic susceptibility is calculated for
$J=\infty$. No ferromagnetic transition is found for any $n$ or $S$ and for
$n=1$ the correct Curie law, with spin $S+1/2$, is found only at high
temperature. On the other hand in \sref{sDMFT} it is shown that for
$S=\infty$, where dynamical mean field theory has been implemented
\cite{rFurukawa}, our CPA equations for the Green functions  agree with
DMFT. Furthermore DMFT leads to a ferromagnetic transition for $0<n<1$ and
to a correct Curie law for $n=1$. In \sref{sDMFTTc} this paradox is
resolved by abandoning the apparently unreliable self-consistent
determination of $\las\exp(\lambda S^z)\ras$ and using instead a
probability distribution $P(S^z)$ to evaluate the required expectation
values. The form of $P(S^z)$ used for finite $S$ is a reasonable extension
of the form which arises in DMFT for $S=\infty$. We then find a finite
Curie temperature $T_{\rm C}$ for $0<n<1$, and correct Curie laws for
$n=0$ and 1, for all $S$. Naturally the results agree with DMFT for
$S=\infty$. The maximum in $T_{\rm C}$, as a function of band-filling $n$,
moves from $n=0.5$ for $S=\infty$ to larger values of $n$ as $S$
decreases.

This work completes our present study of the paramagnetic state and
ferromagnetic transition of the DE model within our many-body CPA. With
some effort we could pursue the calculations into the ferromagnetic state.
However this has already been done for $S=\infty$ within DMFT
\cite{rFurukawa} and the rewards might be slight, particularly since for
finite $S$ the CPA never gives a ground state of complete spin alignment.
It seems more profitable to repair some defects of the DE model itself.
One should include both coupling to phonons and the double degeneracy
of the $e_g$ band. It is likely, as originally proposed by Millis \etal
\cite{rMillis1}, that phonon coupling is essential for an understanding
of the insulator-like paramagnetic state in the manganites. We showed
\cite{rEdwardsGreenKubo} that, without phonons, the DE model gives much
too small a resistivity. The introduction of phonons is therefore a high
priority and it is in fact easier to include coupling to local phonons
in our CPA approach than to consider degenerate orbitals. This is our
next objective.

\ack
We are grateful to K Kubo for helpful discussions in the early stages of
this work.
ACMG was supported by the UK Engineering and Physical Sciences Research
Council (EPSRC).

\end{document}